**Electrically and Magnetically Induced Optical Rotation in $Pb_5Ge_3O_{11}$:Cr Crystals at the Phase Transition. 1. Electrogyration Effect in $Pb_5Ge_3O_{11}$:Cr**

Adamenko D. [1], Klymiv I. [1], Duda V. M. [2], Vlokh R. [1] and Vlokh O. [1]

[1]Institute of Physical Optics, 23 Dragomanov St., 79005 Lviv, Ukraine, e-mail: vlokh@ifo.lviv.ua
[2]Dnipropetrovsk National University, 13 Naukova St., Dnipropetrovsk, Ukraine

**Abstract**

We present the results of studies for temperature dependences of electrogyration (EG) effect and natural optical activity in $Pb_5Ge_3O_{11}$:Cr crystals. Extremely high magnitude of EG coefficient is found for these crystals. We also demonstrate how the Curie-Weiss constant, the critical exponents of the order parameter and the dielectric permittivity, as well as the coefficients of thermodynamic potential, could be derived from the temperature dependences of optical activity and EG coefficient.

**Introduction**

Electrogyration (EG) effect, which consists, in particular, in inducing of optical rotation by a biasing field, has been studied for many crystals (see, e.g., [1]). Among those materials, one can mention the quartz crystals [2,3], Rochelle salt [4], TGS [5] and KDP families [6,7], alums [8], etc. The effect is quite small in dielectric compounds, while the EG coefficients for the wide-band semiconductors such as $Bi_{12}GeO_{20}$ (the coefficients of EG range from $\gamma_{41} = 0.37 \pm 0.03$ to $\gamma_{41}$=$0.05 \pm 0.01$pm/V for the wavelengths $\lambda$=460-760nm [9]), $Bi_{12}TiO_{20}$, $Bi_{12}SiO_{20}$ ($\gamma_{41} = (0.5-3.75)\times 10^{-13} m/V$ [10]) and $AgGaS_2$ ($\gamma_{41} = 2.03\times 10^{-12} m/V$ at $\lambda = 0.498 nm$ [11]) reach larger values, since the dispersion is strong and the bandgap wavelength is located close to the visible spectral range. The other factor that can lead to rise of the EG coefficient is a closeness of proper ferroelectric phase transition, in the vicinity of which EG effect manifest anomalous temperature dependence, due to Curie-Weiss law.

Probably, one of the largest EG rotations has been observed in the lead germanate (LG) crystals close to the Curie temperature [12]. The effect is so large that it has been clearly visible even if one observes the changes in the conoscopic fringes. Of course, any practical utilization of crystals that stay in non-equilibrium conditions, e.g., near the temperatures of phase transitions, is difficult because of a necessity of precise temperature stabilization. However, the temperature of phase transition could be smeared or matched to the normal conditions due to substituting the corresponding chemical elements. Such the behaviour of EG has been observed in the solid solutions $Pb_5Ge_{3x}Si_{3(1-x)}O_{11}$, $(Pb_{1-x}Bi_x)_5Ge_3O_{11}$ and $(Pb_{1-x}Ba_x)_5Ge_3O_{11}$ [13-15]. Notice that the substitution of Pb by Bi in $(Pb_{1-x}Bi_x)_5Ge_3O_{11}$ and the substitution of Ge by Si in $Pb_5Ge_{3x}Si_{3(1-x)}O_{11}$ crystals provides a larger EG peak, when compare with the pure LG crystals. Nonetheless, the opposite situation occurs in $(Pb_{1-x}Ba_x)_5Ge_3O_{11}$ under the substitution of Pb by Ba: the EG coefficient decreases. Another common future appearing at the atomic substitutions in the LG is essential shift in the phase transition point, in comparison with the pure $Pb_5Ge_3O_{11}$ crystals ($T_c = 450$ K – see, e.g., [16]).

Thus, searching for the solid solutions (or doped crystals) with high values of the optical EG rotation is still interesting from the viewpoint of applications of new EG materials for operating optical radiation. Let us stress here that not only chemical substitution could be suspected as promising for this aim, but also doping of pure crystals by some chemical elements. For example, the EG effect has been studied for the LG crystals doped with Li, La [17], Eu [17, 18] and Cd, Nd [19]. The phase transition temperature for $Pb_5Ge_3O_{11}$:Nd crystals is equal to 450 K [20]. Nevertheless, doping of the LG crystals with different chemical elements can lead to shifts in the phase transition temperature, too. Doping of LG crystals is useful for applications in the optical storage, lasing and fabrication of thin ferroelectric film memory units. The photorefractive effect has been studied for Cu and Nd doped LG crystal [21,22]. Concerning the EG in the doped LG crystals, one can remind that the magnitude of EG coefficient in $Pb_5Ge_3O_{11}$:Nd is by an order of magnitude larger than in the pure crystals [17]. In the present work we continue to study EG effect in the doped LG crystals, with a specific emphasis on the material doped with Cr.

**Experimental**

Pure LG crystals undergo a second-order phase transition with the change of point symmetry group $\bar{6} \leftrightarrow 3$. We used the crystal of $Pb_5Ge_3O_{11}$ doped with 0.8 weight % of Cr ions. The samples had the shape of plates, with the thickness $5.0-6.0$ mm and the faces perpendicular to the optic axis. The flatness of the samples was not worse than 0.15 deg. The mismatch of the orientation of faces with respect to the direction perpendicular to the optic axis was less than 0.32 deg. The electric fields up to $10^6$ V/m were applied to the crystals along the optic axis direction with the aid of transparent electrodes (the glass plates coated by conducting tin oxide layer). Thin glass plates with the thickness of $d = 0.15$ mm have been introduced between the sample and transparent electrodes for avoiding the appearance of electric current. Thus, the electric voltage between the sample surfaces was lower by 5%, when compare with that applied to the electrodes.

The experimental set-up is presented in Fig. 1. The sample was placed into a furnace enabling temperature stabilization and control with the accuracy $\pm 0.1\mathrm{K}$. The optical activity was measured after determining orientation of polarization plane of the light emerged from the sample, for the known linear polarization state of the incident light propagated along the optic axis of crystal. Faraday modulator was used for increasing precision of measurements of the optical rotation.

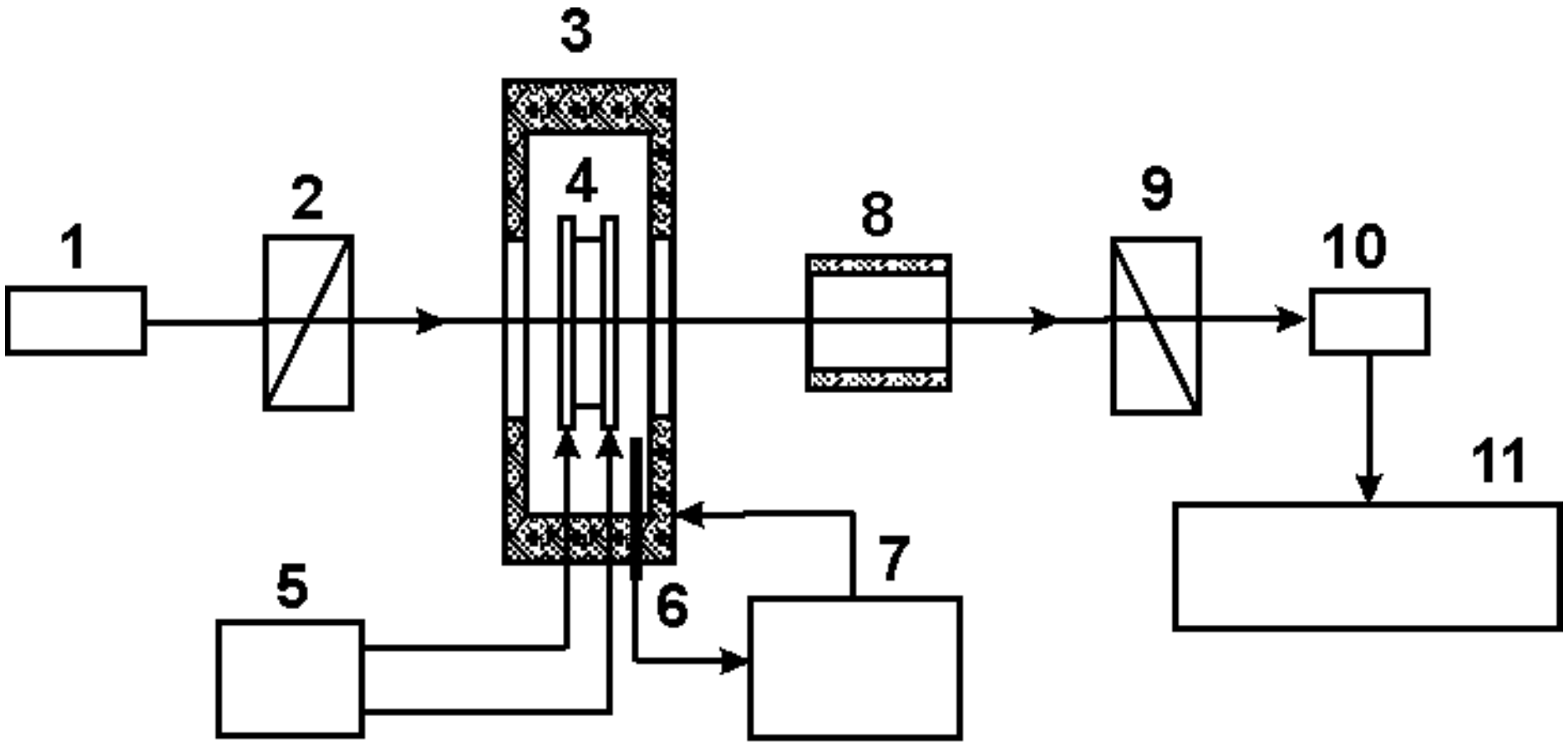

Fig. 1. Experimental set-up: 1 – He-Ne laser, 2, 9 – polarizers with rotation stages, 3 – furnace, 4 – sample with transparent electrodes, 5 – high voltage supply, 6 – thermocouple, 7 – temperature controller, 8 – Faraday cell, 10 – photo multiplier, 11 – oscilloscope.

Though the accuracy for the measurements of polarization orientation was not worse than $0.1\deg$, the total errors were higher. For example, the magnitude of the optical activity $\rho$ during the measurements of temperature dependences of the natural optical rotation was determined with the error of 7% because of slight unipolarity of crystals; the electrically induced increment of the optical rotatory power $\Delta\rho$ was measured with the error 2–10% at the temperatures lower than $\sim 460\mathrm{K}$ and the mentioned error still increased at higher temperatures up to 10–25%, since high temperatures caused a damage of film electrodes; the EG coefficient was determined with the same error as the increment of optical rotatory power. When measuring the temperature dependences of optical activity, we applied the biasing field in the paraelectric phase along $\pm z$ direction and then cooled the sample down to the ferroelectric phase, in order to reach a single-domain state.

The EG effect is described by the relations

$$\varepsilon_{ij} = \varepsilon_{ij}^0 + ie_{ijk} g_{kl} k_l = \varepsilon_{ij}^0 + ie_{ijk} (g_{kl}^0 + \gamma_{klm} E_m + \beta_{klmn} E_m E_n) k_l, \qquad (1)$$

$$\Delta g_{kl} = g_{kl} - g_{kl}^0 = \gamma_{klm} E_m + \beta_{klmn} E_m E_n , \qquad (2)$$

or

$$\Delta g_{kl} = \tilde{\gamma}_{klm} {}^{(s)}P_m + \tilde{\beta}^{(s)} P_m {}^{(s)}P_n , \qquad (3)$$

$$\rho = \frac{\pi}{\lambda n} G = \frac{\pi}{\lambda n} g_{kl} l_k l_l , \qquad (4)$$

in terms of spontaneous polarization. Here $\varepsilon_{ij}$ and $\varepsilon_{ij}^0$ are the optical-frequency dielectric permittivities that take into account and disregard the spatial dispersion phenomena, respectively; $e_{ijk}$ the Levi-Civita tensor; $g_{kl}$ and $g_{kl}^0$ the second-rank axial tensors that describe total and natural optical activity, respectively; $k_l$ the wave vector of light; $E_m$, $E_n$ and ${}^{(s)}P_n$, ${}^{(s)}P_m$ respectively the components of the electric field and the spontaneous polarization; $\gamma_{klm}$, $\beta_{klmn}$ and $\tilde{\gamma}_{klm}$, $\tilde{\beta}_{klmn}$ the third- and forth-rank axial tensors describing the linear and quadratic EG effects expressed in terms of electric field and spontaneous polarization, respectively; $G$ the pseudoscalar gyration parameter; and the components of the unit wave vector in the spherical coordinate system; $\lambda$ the light wavelength; and $n$ the refractive index for the given propagation direction. The matrices of the tensors of linear EG effect for the point groups of symmetry $\bar{6}$ and $3$ are respectively as follows:

$$\gamma_{klm} = \begin{array}{c|ccc} & E_1 & E_2 & E_3 \\ \hline \Delta g_1 & 0 & 0 & \gamma_{13} \\ \Delta g_2 & 0 & 0 & \gamma_{13} \\ \Delta g_3 & 0 & 0 & \gamma_{33} \\ \Delta g_4 & \gamma_{41} & \gamma_{42} & 0 \\ \Delta g_5 & \gamma_{42} & -\gamma_{41} & 0 \\ \Delta g_6 & 0 & 0 & 0 \end{array} , \quad \gamma_{klm} = \begin{array}{c|ccc} & E_1 & E_2 & E_3 \\ \hline \Delta g_1 & \gamma_{11} & \gamma_{12} & \gamma_{13} \\ \Delta g_2 & -\gamma_{11} & -\gamma_{12} & \gamma_{13} \\ \Delta g_3 & 0 & 0 & \gamma_{33} \\ \Delta g_4 & \gamma_{41} & \gamma_{42} & 0 \\ \Delta g_5 & \gamma_{42} & -\gamma_{41} & 0 \\ \Delta g_6 & -\gamma_{12} & -\gamma_{11} & 0 \end{array} . \qquad (5)$$

It follows from the form of these tensors that the linear EG effect really exists for the experimental geometry $k \parallel E \parallel z$ chosen by us. In this case the optical rotatory power due to the EG may be presented as

$$\rho = \frac{\pi}{\lambda n_o} \gamma_{33} E_3 , \qquad (6)$$

with $n_o$ meaning the ordinary refractive index. Since the so-called second-order symmetry operations are lost at the phase transition with the symmetry change $\bar{6} \to 3$, the domains in the ferroelectric phase should be enantiomorphous. This should lead to reversal of the natural optical activity whenever the domain structure is switched over by means of the biasing field $E_3$. At the same time, the LG crystals should possess no natural optical activity in the paraelectric phase, due to general symmetry limitations.

**Results and discussion**

The gyration hysteresis loops obtained at different temperatures close to the Curie point are shown in Fig. 2. One can see that the hysteresis loops become narrower when approaching $T_c$, due to decreasing coercive field. Moreover, the slopes of the linear field dependences of the optical rotation power out of the hysteresis loops then increase. This indicates indirectly that the EG coefficient $\gamma_{33}$ increases when the temperature increases up to $454\,\mathrm{K}$.

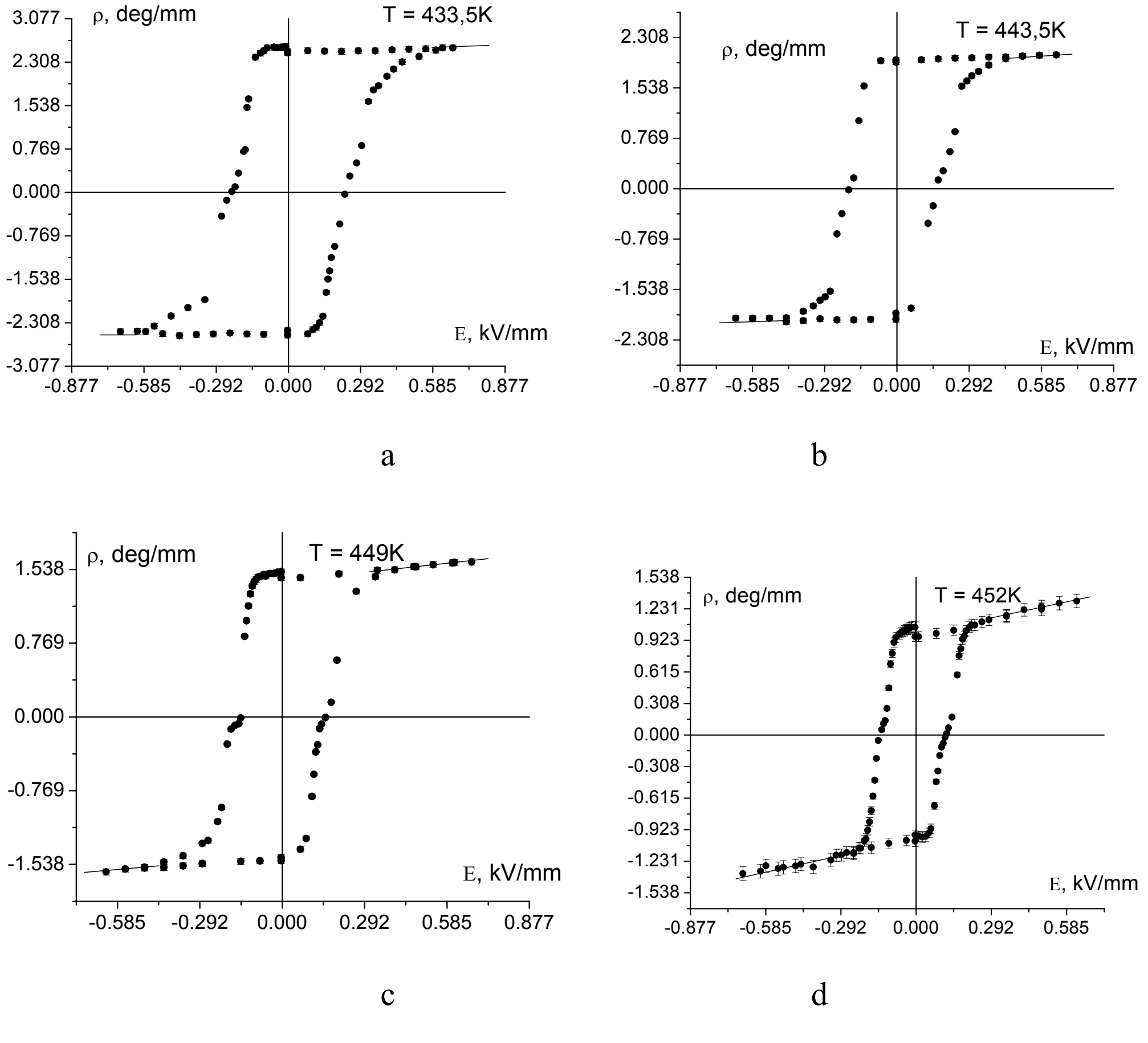

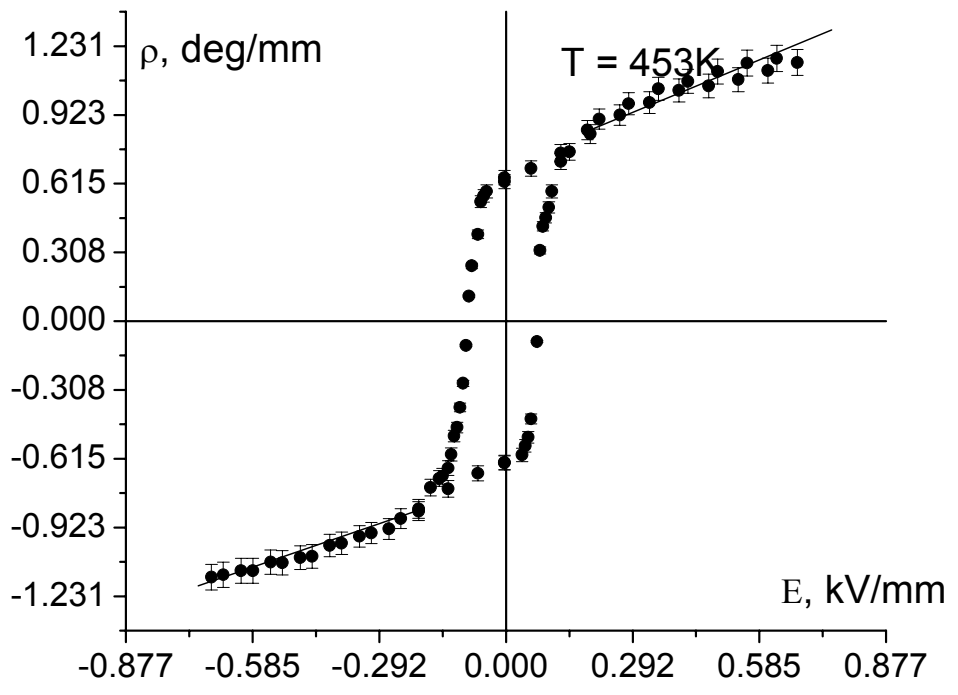

e

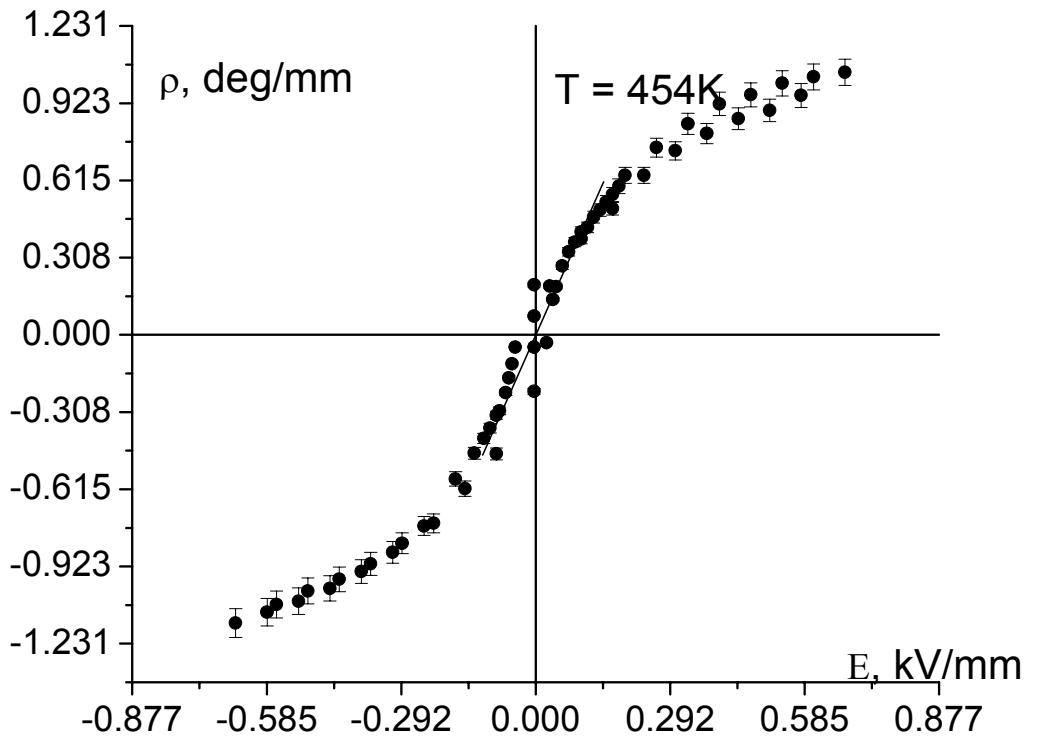

f

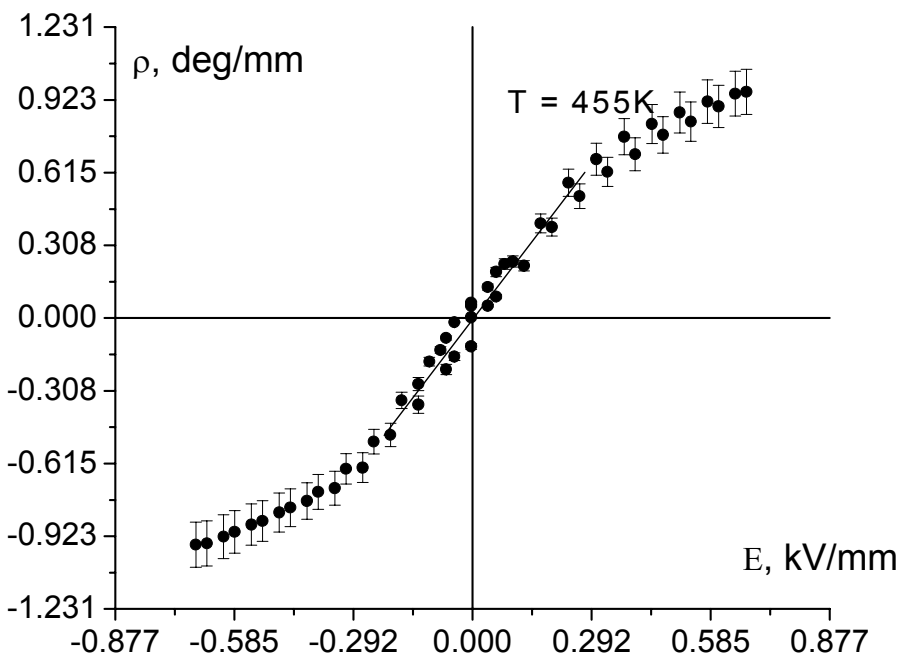

g

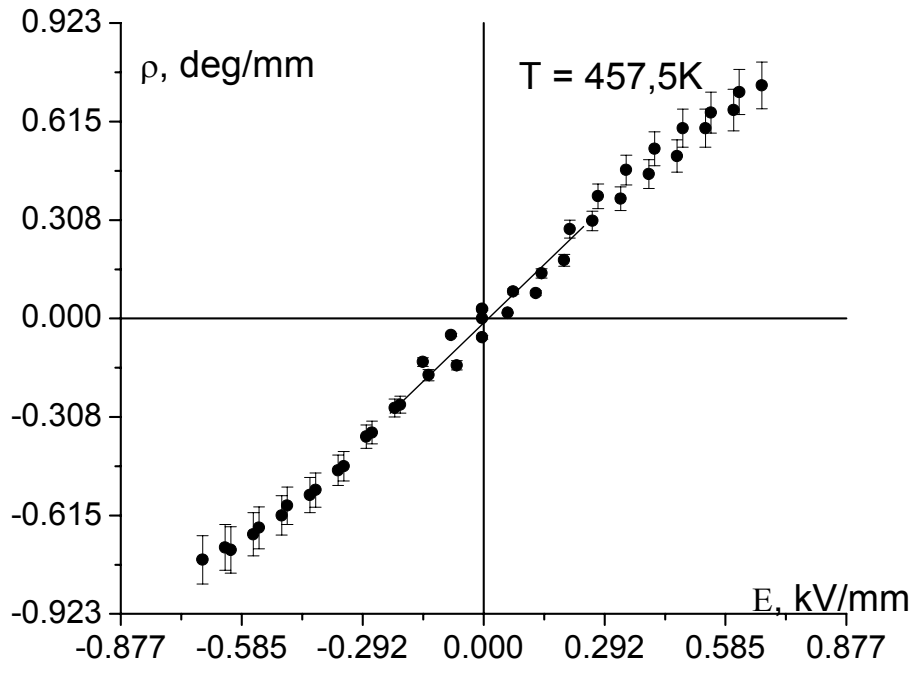

h

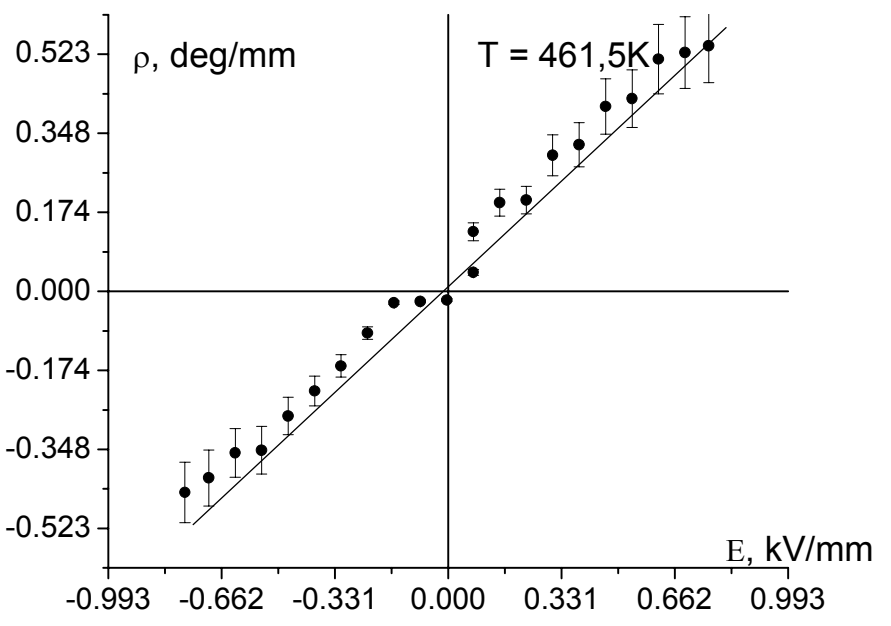

i

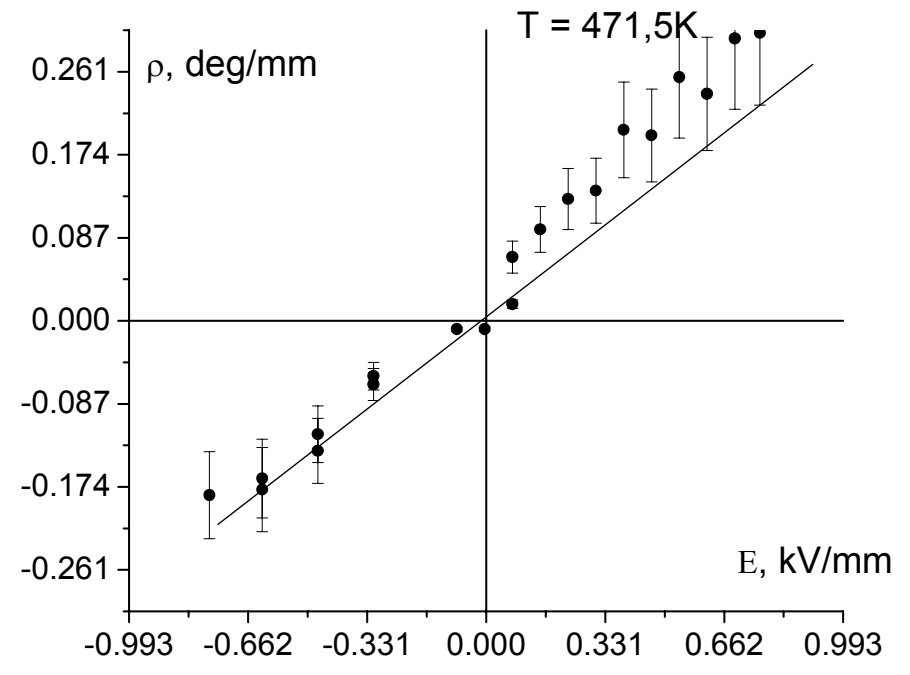

j

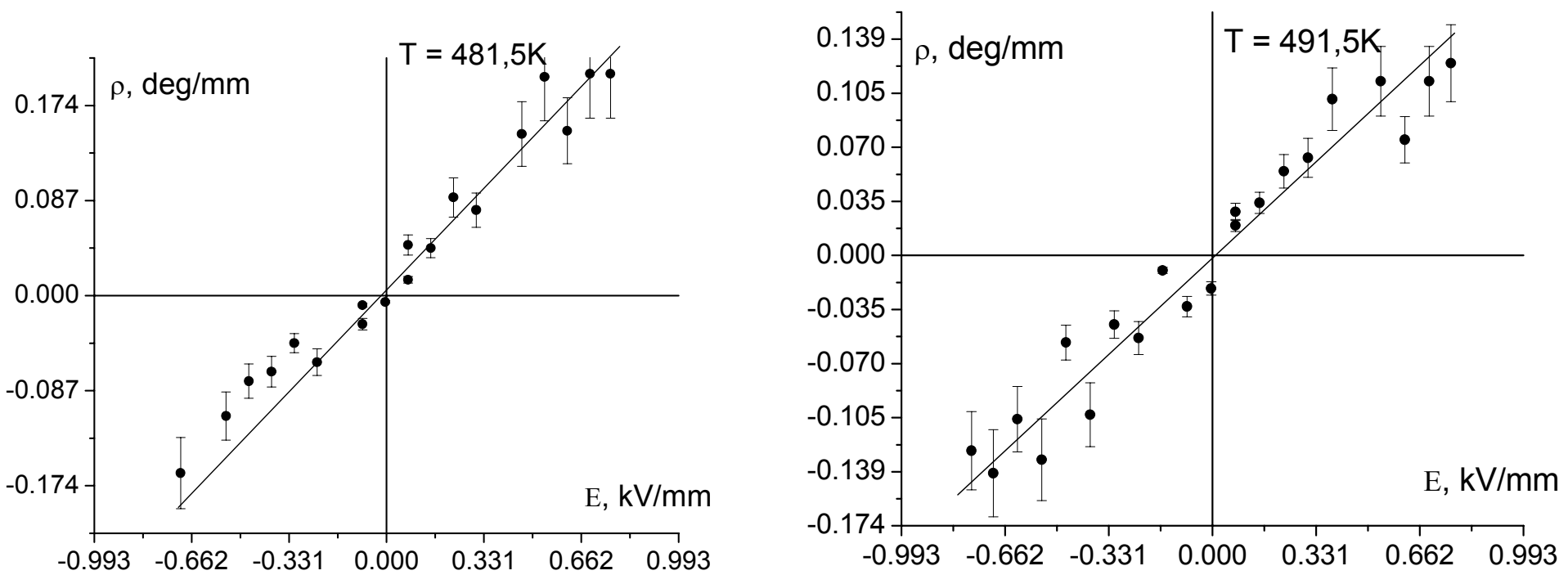

k l

Fig. 2. Dependences of optical rotatory power for $Pb_5Ge_3O_{11}$:Cr crystals on the biasing field $E_3$ at different temperatures ( $\lambda = 632.8\,\text{nm}$ ). Bars show the relative errors and straight lines indicate the optical rotation increments associated with the EG.

One can see (Fig. 2f, g) that the field dependences of the optical rotation still remain nonlinear at the temperatures 454 K and 455 K, while the hysteresis loop is shifted from the origin of coordinates up to $(0.3-0.7)\,\text{kV/mm}$. Obviously, such the behaviour could be explained by the fact that the application of biasing field parallel or anti-parallel to the spontaneous polarization in the vicinity of $T_c$ leads to additional smearing of the phase transition, shifting of $T_c$ and inducing of ferroelectric phase and micro-domains in all the sample volume. Probably, the electric field $0.3\,\text{kV/mm}$ at $T = 454\,\text{K}$ is sufficient to induce ferroelectric phase in the sample. Only when the electric field increases above this value, does the domain structure begin to switch. Thus, we determine the temperature $T = 454\text{K}$ (in the cooling regime) as a finishing point of nucleation of feroelectric phase in the matrix of paraelectric phase of $Pb_5Ge_3O_{11}$:Cr crystals. This temperature is only $4\,\text{K}$ higher than the corresponding phase transition temperature for the pure LG crystal. If we take into consideration a diffuse character of the second-order phase transition in Cr-doped crystals (see Fig. 3), we arrive at the conclusion that the Curie temperature for the $Pb_5Ge_3O_{11}$:Cr crystals does not essentially differ from that of the pure LG. It also follows from Fig. 3 that the processes related to the phase transition begin at $T = 457\,\text{K}$ in the cooling run (the temperature at which the optical activity appears) and they are completed at $T = 454\,\text{K}$ . Thus, the temperature region of $454\,\text{K} - 457\,\text{K}$ might be defined as a region, in which the phase transition is smeared. As one can easily see (Fig. 2h), the dependence of optical activity on the biasing field is close to linear at $457.5\,\text{K}$ , thus implying that the crystals under study are in the paraelectric phase at this temperature. It is worth mentioning that the linear dependences of optical rotation on the biasing field have been observed in the paraelectric phase up to 491.5 K (see Fig. 2i–l).

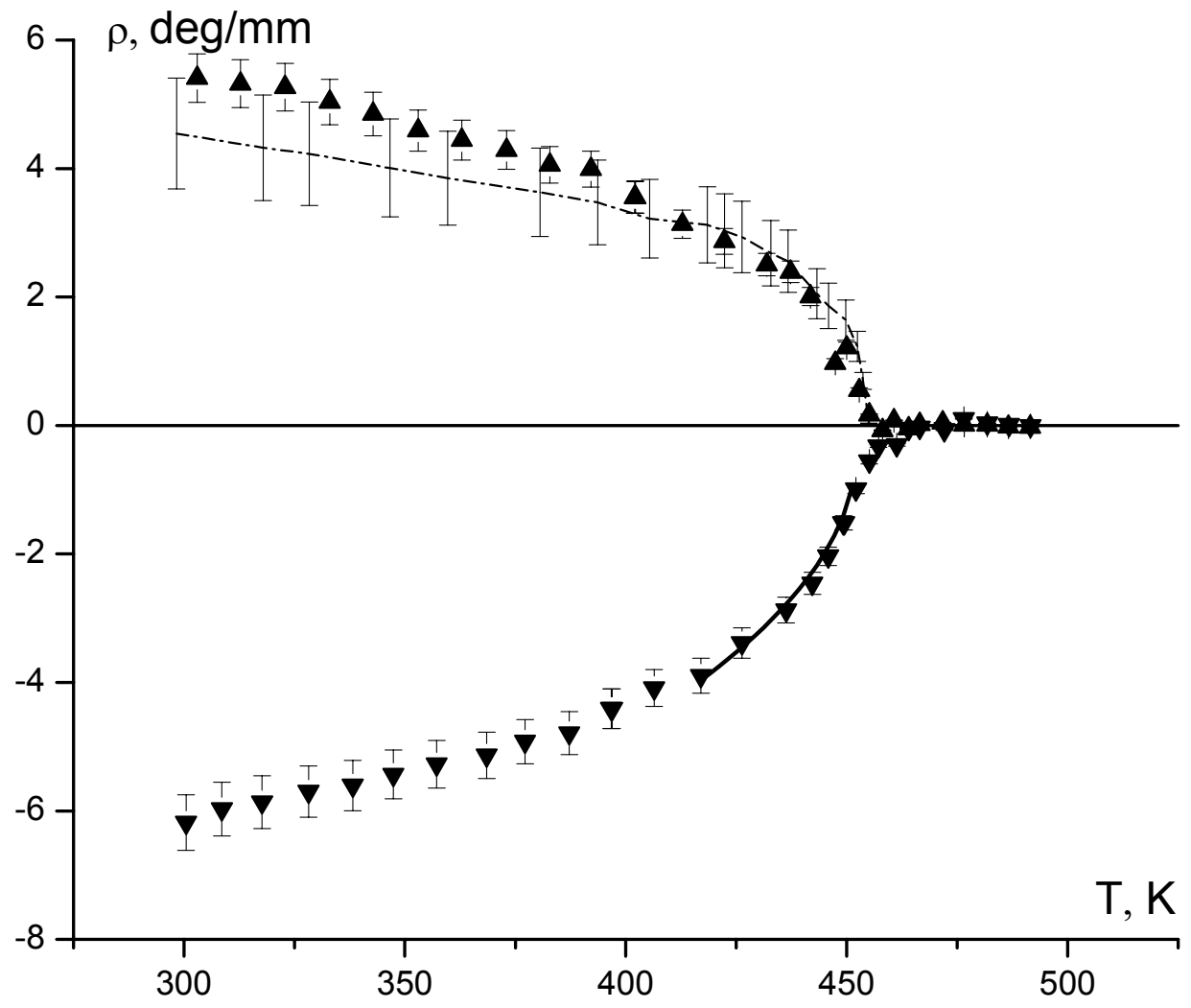

Fig. 3. Temperature dependences of optical rotatory power for the opposite single-domain states (full triangles) in $Pb_5Ge_3O_{11}$:Cr crystals ($\lambda = 632.8$ nm; bars indicate the relative errors). Dashed line corresponds to the temperature dependence of optical rotatory power calculated with the spontaneous polarization data for the pure $Pb_5Ge_3O_{11}$ [16] and the relation $\rho = \frac{\pi}{\lambda n_o} \tilde{\gamma}_{klm}{}^{(s)} P_m$ (bars indicate the relative errors). Solid line corresponds to fitting of the temperature dependence, using the critical exponent $\mu = 0.5$.

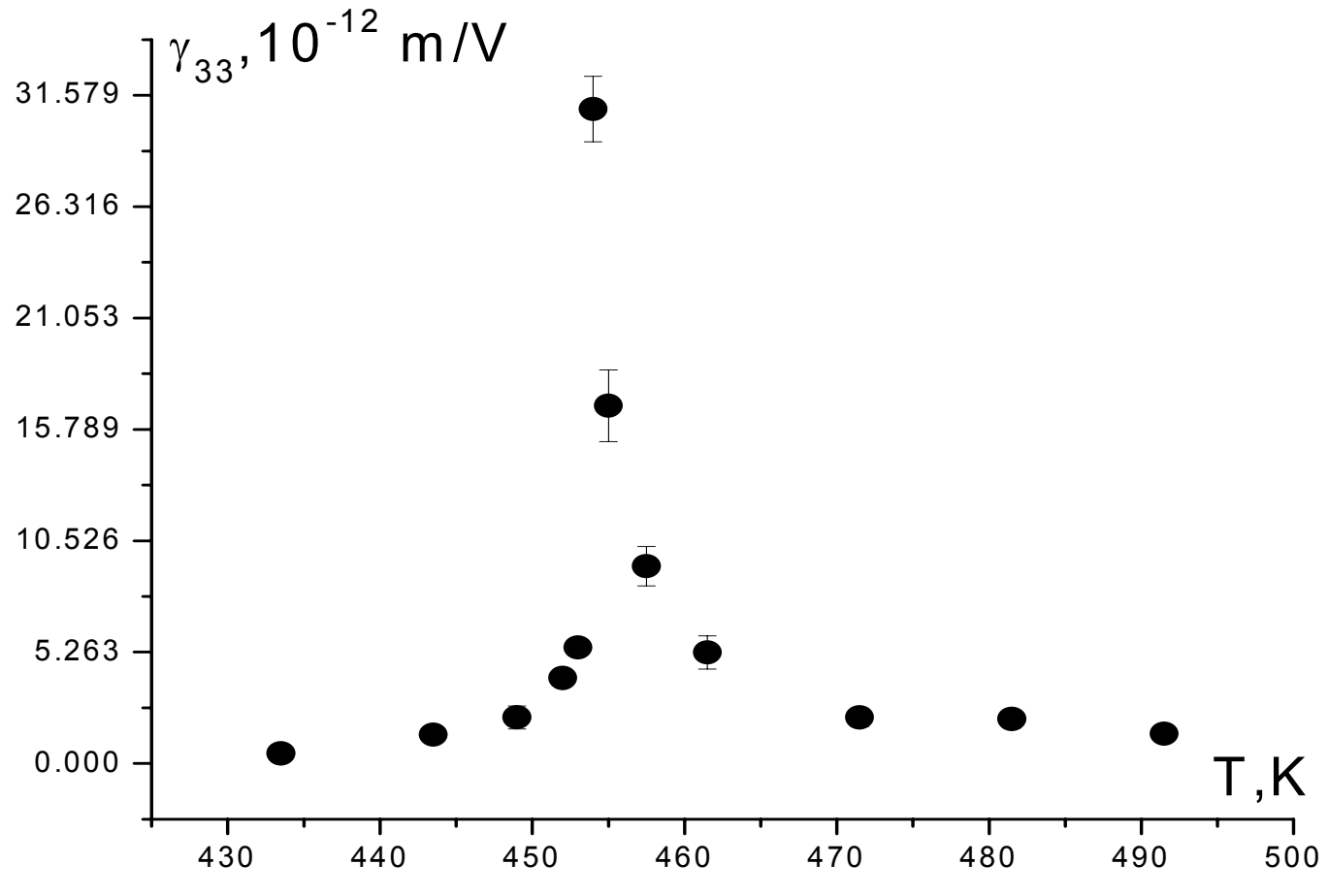

a

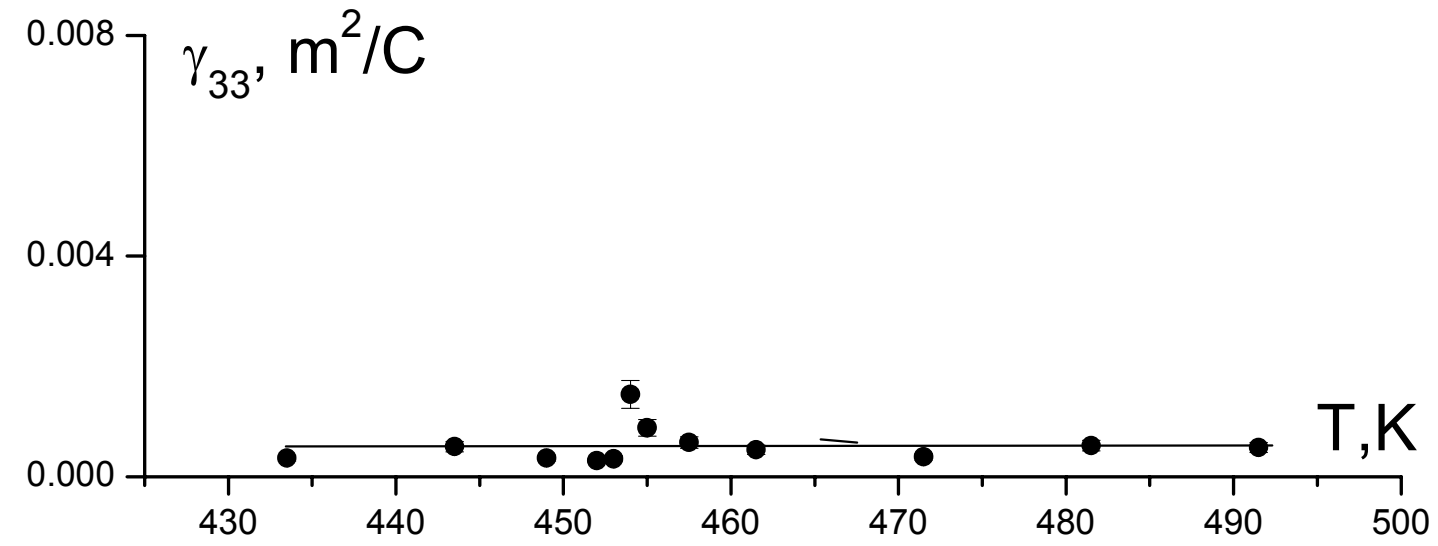

b

Fig. 4. Temperature dependences of EG coefficient $\gamma_{33}$ for $Pb_5Ge_3O_{11}$:Cr crystals obtained experimentally (a) and the coefficient $\tilde{\gamma}_{33}$ (b) calculated with the relation $\tilde{\gamma}_{33} \simeq \frac{\gamma_{33}}{\varepsilon_0 \varepsilon_3}$ ($\lambda = 632.8$ nm). Bars indicate the relative errors.

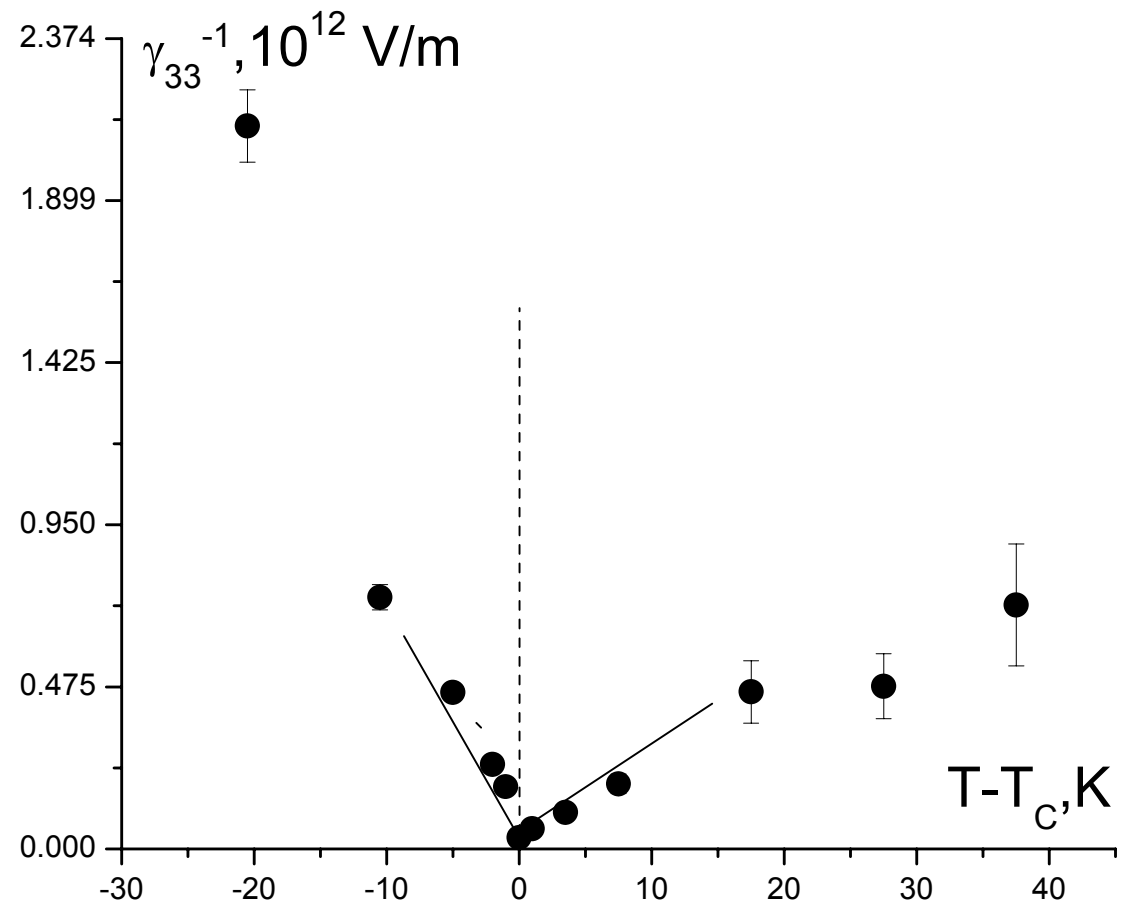

Fig. 5. Dependence of reciprocal EG coefficient on $T - T_c$ for Cr-doped $Pb_5Ge_3O_{11}$ crystals (bars indicate the relative errors).

The natural optical activity for the Cr-doped $Pb_5Ge_3O_{11}$ crystals at room temperature is approximately the same as in the pure LG. The magnitudes of the optical rotatory power for the opposite single-domain states are $5.4$ deg/mm and $6.2$ deg/mm at $T = 300$ K (see Fig. 3). The difference of those values should be explained by unipolarity of our samples. The mean value of the optical rotatory power for the Cr-doped $Pb_5Ge_3O_{11}$ crystals is therefore equal to $5.8 \pm 0.4$ deg/mm at $T = 300$ K, while, e.g., for the pure $Pb_5Ge_3O_{11}$ it is $\sim 5.9$deg/mm [1,2].

Let us now consider thermodynamic potential for the proper, second-order phase transition (the terminology see, e.g., in [23]) observed in the LG-type crystals:

$$F = F_0 + \alpha(T - T_c)\,^{(s)}P^2 + \frac{1}{2}\beta\,^{(s)}P^4 + \frac{1}{3}\delta\,^{(s)}P^6 + \dots \quad (7)$$

After accounting for Eq. (7) and the minimization conditions

$$\left(\frac{\partial F}{\partial\,^{(s)}P}\right) = 0, \quad \left(\frac{\partial^2 F}{\partial\,^{(s)}P^2}\right) > 0, \quad (8)$$

we obtain the solution

$$^{(s)}P^2 = \frac{\beta}{2\delta}\left(-1 \pm \sqrt{1 - \frac{4\alpha(T - T_c)\delta}{\beta^2}}\right) \text{ at } T < T_c. \quad (9)$$

In the SI units we have close to $T_c$

$$^{(s)}P^2 = -\frac{8\pi\alpha(T - T_c)}{\beta}. \quad (10)$$

Let us take Eq. (10) and the relations $g_3 = \gamma_{33}\,^{(s)}P_3$ and $\rho = \frac{\pi}{\lambda n_o} g_3$ into account. Then we get

$$\rho = \pm\frac{2\tilde{\gamma}_{33}}{\lambda n_o}\left[\frac{2\pi\alpha}{\beta}(T_c - T)\right]^{1/2}. \quad (11)$$

One can see (Fig. 3) that the dependence of optical rotation close to $T_c$ is well fitted by the function $(T_c - T)^{\mu}$, where $\mu = 0.5$. Thus, Eq. (11) is satisfied for $Pb_5Ge_3O_{11}$:Cr crystals in the temperature region $T_c - 40\text{ K} < T < T_c - 5\text{ K}$. Notice that the critical exponent for the pure LG crystals has been found to be equal to 0.35 for the temperature interval between the room temperature and $T = T_c - 3\text{ K}$ [24]. It is interesting that we observe a tail of optical activity above $T_c - 5\text{ K}$ and the dependence of optical activity on $(T_c - T)^{0.5}$ still remains linear below $T < T_c - 40\text{ K}$, though the slope is then changed. A similar behaviour has been earlier found for the square of spontaneous polarization in the pure LG crystals [16].

In order to calculate EG coefficient (see Eq. (6)), we have used the data presented in Fig. 2 and the value of refractive index $n_o = 2.12$ for the pure LG crystals [25]. This is reasonable, since the properties of pure and Cr-doped compounds (e.g., the values of the natural optical activity and the phase transition temperatures) are quite close. It is seen from Fig. 4a that the EG coefficient $\gamma_{33}$ manifests anomalous behaviour in the vicinity of $T_c$. It reaches the value $\gamma_{33} = (3.1 \pm 0.3) \times 10^{-11}\text{m/V}$ at $T_c$. This is the largest coefficient measured so far for the LG-type crystals and, maybe, the highest EG magnitude ever achieved. Basing on the thermodynamic potential defined by Eq. (7), with the additional term ($-PE$), and the condition $\varepsilon = \frac{\partial D}{\partial E}$, one readily obtains for the behaviour of dielectric permittivity:

$$\varepsilon = \frac{C}{T - T_c} \text{ at } T > T_c \tag{12}$$

and

$$\varepsilon = \frac{C}{2(T_c - T)} \text{ at } T < T_c, \tag{13}$$

where $C = (2\alpha)^{-1}$ is the Curie-Weiss constant. Considering the relation $\gamma_{33} = \tilde{\gamma}_{33}\varepsilon_0(\varepsilon_3 - 1) \simeq \tilde{\gamma}_{33}\varepsilon_0\varepsilon_3$ (with $\varepsilon_0$ being the permittivity of free space), we have

$$\gamma_{33} = \frac{C\varepsilon_0\tilde{\gamma}_{33}}{T - T_c} \text{ for } T > T_c, \tag{14}$$

and

$$\gamma_{33} = \frac{C\varepsilon_0\tilde{\gamma}_{33}}{2(T_c - T)} \text{ for } T < T_c. \tag{15}$$

Hence,

$$\frac{(\gamma_{33}^{-1})_{T<T_c}}{(\gamma_{33}^{-1})_{T>T_c}} = 2. \tag{16}$$

As seen from Fig. 5, Eq. (16) is approximately satisfied in the temperature range close to $T_c$ (the corresponding ratio is equal to $\sim 2.4$, close to the value 3.0 found for the pure LG crystals [16]). When calculating the EG coefficient $\tilde{\gamma}_{33}$, we have used the temperature dependence of the dielectric permittivity for the pure LG crystals presented in [16]. It is clear (see Fig. 4b) that the recalculated coefficient $\tilde{\gamma}_{33} \simeq \frac{\gamma_{33}}{\varepsilon_0\varepsilon_3}$ does not depend on temperature and is equal to $(6.0 \pm 1.0)\times 10^{-4}$ m$^2$/C. Using this value, the relation $\rho = \frac{\pi}{\lambda n_o}\tilde{\gamma}_{klm}{}^{(s)}P_m$ and the temperature dependence of spontaneous polarization for the pure LG crystal [16], one can estimate the temperature dependence of the optical rotatory power (see Fig. 3). The dependence obtained experimentally is in a good agreement with the theoretical one. This means that the phase transition in the Cr-doped LG crystals is indeed a proper ferroelectric one and the changes in the physical properties in the course of this transition (the optical activity in our case) are described in terms of the appearance of spontaneous electrical polarization. It is interesting to note that the Curie-Weiss constant for the Cr-doped LG crystals calculated on the basis of temperature dependence of the EG coefficient is equal to $C = (0.4 \pm 0.1)\times 10^4$ K, while for the pure crystals it is $C \approx 1.04\times 10^4$ K [16]. Finally, we have calculated the coefficients of thermodynamic potential ($\alpha \simeq 1.25\times 10^{-4}$ K$^{-1}$ and $\beta \approx 5\, m^4/C^2$), using the temperate dependence of optical rotatory power (Fig. 3), Eq. (11) and the Curie-Weiss constant.

**Conclusions**

In the present paper we have studied EG effect in the Cr-doped LG crystals, induced by external biasing field and spontaneous polarization in the vicinity of phase transition. It has been found that the EG coefficient reaches extremely large value ($\gamma_{33} = (3.1 \pm 0.3) \times 10^{-11}$ m/V at $T_c$). Probably, the latter is the highest among the EG coefficients for all of the known materials. On the basis of temperature dependences of the optical rotatory power and the EG coefficient it is shown that the phase transition in $Pb_5Ge_3O_{11}$:Cr crystals is a proper second-order ferroelectric one. The Curie-Weiss constant, critical exponent for the order parameter, the dielectric permittivity and the coefficients of thermodynamic potential have been calculated following from the temperature dependences of optical activity and EG coefficient.

**Acknowledgement**

The authors acknowledge financial support of this study from the Ministry of Education and Science of Ukraine (the Project N0106U000616).